\documentstyle[aps,prl,epsf,floats]{revtex}
\begin{document}
\twocolumn[\hsize\textwidth\columnwidth\hsize\csname@twocolumnfalse%
\endcsname
\title{Comment on ``Critical spin dynamics of the 2D quantum Heisenberg
antiferromagnets: Sr$_2$CuO$_2$Cl$_2$ and
Sr$_2$Cu$_3$O$_4$Cl$_2$''}
\author{Subir Sachdev$^{1,2}$ and Oleg A. Starykh$^{3}$}
\address{$^{1}$Department of Physics, Harvard University,
Cambridge MA 02138,\\ $^{2}$Department of Physics, Yale
University, P.O. Box 208120, New Haven, CT 06520-8120\\
$^{3}$Department of Physics, Hofstra University, Hempstead, NY
11549}
\date{January 22, 2001}
\maketitle
\begin{abstract}
We compare the neutron measurements of Kim {\em et al.}
(cond-mat/0012239) on two-dimensional, $S=1/2$ antiferromagnets
with the continuum quasiclassical theory of S. Sachdev and O.A.
Starykh (Nature, {\bf 405}, 322 (2000)). The damping of the lowest
energy spin excitations is characterized by a dimensionless number
whose temperature dependence was predicted to be determined
entirely by that of the uniform spin susceptibility. Theory and
experiment are consistent with each other.
\end{abstract}
\pacs{PACS numbers:}
]
% \newpage

Kim {\em et al.} \cite{yjkim} have recently provided illuminating
neutron scattering measurements of the damping of the lowest
energy spin excitations in $S=1/2$, square lattice
antiferromagnets. They characterized this damping by the
temperature ($T$) dependence of a dimensionless number,
$R_{\omega}$. (Related measurements, at higher energies, are
those of \cite{ronnow}.) Here we will use an existing
quasiclassical dynamic theory \cite{oleg} to relate $R_{\omega}
(T)$ to a thermodynamic observable, $\chi_u (T)$, the uniform
spin susceptibility, and show that this provides a unified
understanding of the experiments of Kim {\em et al.}, and earlier
studies of NMR relaxation rates \cite{tognetti}.

Detailed quantitative predictions have been made for $\chi_u (T)$
(and other static observables) for antiferromagnets with a
magnetically ordered ground states with a spin stiffness $\rho_s$
\cite{CHN,CSY} . When the antiferromagnet is also near a quantum
critical point at which $\rho_s$ vanishes, then these observables
become universal functions of $T/\rho_s$ at all $T$ below $J$ (a
near-neighbor exchange interactions), while at higher $T$ there is
simple decoupled spin behavior. In particular, $\chi_u$ obeys
\cite{CSY} ($k_B = \hbar = 1$)
\begin{equation}
\chi_u (T) = (T/c^2) \Omega(T/\rho_s), \label{e1}
\end{equation}
where $c$ is the spin-wave velocity, and $\Omega(x)$ is a
universal function which crosses over from the `renormalized
classical' (RC) regime $\Omega (x \ll 1) = 2/(3x) + 1/(3
\pi)+\ldots$, to the `quantum critical' (QC) regime
$\Omega(\infty) \approx 0.27$ \cite{CSY}. Such results agree well
with Monte Carlo simulations on antiferromagnets known to be near
a quantum critical point \cite{troyer1}.

For the square lattice antiferromagnet, RC behavior is well
established as $T \rightarrow 0$. However, somewhat surprisingly,
it was found that $\chi_u (T)$ was quite close to the QC limit
above for $T> 0.3J$, and in clear disagreement with the $T$
dependence predicted by the RC limit, suggesting a crossover from
RC to QC behavior even in this unfrustrated, isotropic, and
undoped system\cite{CSY}; the corresponding behavior was not seen
for the correlation length, and a theoretical rationale  was
offered for this\cite{CSY}. Subsequent precision Monte Carlo
studies were consistent with these observations \cite{troyer2}.

It is clearly of interest to obtain quantitative predictions of
the RC to QC crossover in dynamic properties. However, accurate
predictions of damping rates in the QC regime are quite difficult
to obtain. An expansion in $\epsilon=3-d$ ($d$ is the spatial
dimensionality) was developed in \cite{relax}, but the accuracy of
the leading order term in $\epsilon$ is not known. In \cite{oleg},
physical arguments were used to motivate a simple continuum
quasiclassical model which had the advantages of being expressed
directly in $d=2$, and of also describing the crossover into the
RC regime. The scaling arguments of \cite{oleg} predict that
$R_{\omega} (T)$ obeys
\begin{equation}
R_{\omega} (T) = A_{\omega} \sqrt{T/(c^2 \chi_u (T))}, \label{e2}
\end{equation}
where $A_{\omega}$ is a dimensionless number. So in \cite{oleg},
{\em the crossover in the damping is determined entirely by that
in $\chi_u (T)$.} Upon applying a self-consistent, one-loop
approximation to the theory of \cite{oleg}, we obtain integral
equations closely related to those of Grempel \cite{grempel}; from
his numerical solution of these equations, we deduce $A_{\omega} =
0.31$.

In the $T \rightarrow 0$, RC limit, (\ref{e2},\ref{e1}) predict
that $R_{\omega} (T) = 0.38\sqrt{T/\rho_s}$, which is precisely
the result of \cite{CHN,grempel}. In the QC regime, it is a very
general prediction that $R_{\omega}(T)$ is a $T$-independent
constant, as appears to be observed at higher $T$ in \cite{yjkim};
from (\ref{e2},\ref{e1}) we obtain that $R_{\omega} (T) \approx
0.60$, a prediction which is consistent with the observations of
\cite{yjkim}. It would be interesting to use the measured values
of $\chi_u (T)$ to test (\ref{e2}) over the entire temperature
range.

Similar comments apply to the computation of NMR relaxation rates
in \cite{tognetti}. These papers use a computation very similar to
that of Grempel \cite{grempel}, but with a lattice cutoff, and
include the full $T$ dependence of $\chi_u (T)$. As we noted
earlier, the latter is not described by the RC behavior but
displays a RC to QC crossover; in the model of \cite{oleg}, the
crossover in $\chi_u (T)$ is sufficient to describe the crossover
in the dynamics into the QC regime. The agreement of the results
in \cite{tognetti} with experimental observations is therefore
consistent with our discussion here.

This research was supported by NSF Grant DMR 96--23181.

\end{document}